\documentclass[nologo,url,11pt,a4paper]{ETHpaper}

\usepackage{amssymb,amsmath} 
\usepackage{graphicx}
\usepackage[numbers,sort&compress]{natbib} 
\usepackage[utf8]{inputenc}
\usepackage{pict2e}

\newcommand{\abs}[1]{\left| #1 \right|}
\newcommand{\mean}[1]{\left\langle #1 \right\rangle}

\begin{document}

\title{Value of peripheral nodes in controlling multilayer networks}
\titlealternative{The value of peripheral nodes in controlling multilayer networks}
\author{Yan Zhang, Antonios Garas \& Frank Schweitzer}
\authoralternative{Y. Zhang A. Garas \& F. Schweitzer}
\address{Chair of Systems Design, ETH Zurich, Weinbergstrasse 58, 8092
   Zurich, Switzerland}

\reference{Phys. Rev. E 93, 012309 (2016) }

\www{\url{http://www.sg.ethz.ch}}

\makeframing
\maketitle

\begin{abstract}
We analyze the controllability of a two-layer network, where driver nodes can be chosen randomly only from one layer. 
Each layer contains a scale-free network with directed links, and the node dynamics depends on the incoming links from other nodes.  
We combine the in-degree and out-degree values to assign an importance value $w$ to each node, and distinguish between peripheral nodes with low $w$ and central nodes with high $w$.
Based on numerical simulations, we find that, the controllable part of the network is larger when choosing low $w$ nodes to connect the two layers. 
The control is as efficient when peripheral nodes are driver nodes as it is for the case of more central nodes. 
However, if we assume a cost to utilize nodes which is proportional to their overall degree, utilizing peripheral nodes to connect the two layers or to act as driver nodes is not only the most cost-efficient solution, it is also the one that performs best in controlling the two-layer network among the different interconnecting strategies we have tested. 
\end{abstract}

\section{Introduction}
How can we efficiently control the dynamics on complex multilayer networks if we are able to control only a few nodes? 
This problem is of importance in the growing field of interconnected networks \citep{kivela2014,Garas2015}.
Such networks consist of multiple layers that each contain a complex network, and additional links between nodes of different layers. 
In addition to its structural properties, such as degree, each node is characterized by a dynamical variable $x_{i}(t)$ that changes dependent on the interaction with other nodes. 
Hence, we face a combined problem in which the dynamics of $N$ coupled equations, where $N$ is the total number of nodes, is exacerbated by the rather complex coupling between these nodes both through intra-layer and inter-layer links. 
The question then is (a) how many, and (b) which of these nodes we need to control in order to control most of the whole network. 

According to control theory, controllability characterizes the ability to drive a dynamical system from any initial state to any desired final state in finite time, by attaching control signals to a carefully chosen set of driver components. 
In the context of complex networks, fairly recently \citet{Liu2011} developed an analytical framework to study the controllability of single-layer complex networks. 
This assumes a linear dynamics for the nodes. Recent efforts have aimed at understanding the interplay between the topological structure and the controllability of complex networks \citep{Liu2014}.
However, a full control of large-scale complex networks has hardly been achieved. In addition to the sheer size of such systems, there are constraints in accessing all of the driver nodes necessary to control the system~\cite{Gao2014a}.
This limitation motivates our research, namely to understand how this control can be achieved with a rather small number of driver nodes. 
We further want extent the scope from single-layer to multilayer networks and consider, as an additional challenge, restricted access to only one layer of a multilayer network.

This merges two research lines, namely controllability of complex networks and multilayer networks, which were jointly discussed so far in a few publications only \citep{Yuan2014,nie2015effect,Menichetti2015}.
All these works, however, focus on the controllability of the whole system and there is no restriction when choosing driver nodes, which is a main issue addressed by our paper.
More precisely, Yuan et al.\citep{Yuan2014} deployed the exact controllability theory to study controllability of multiplex networks; Nie et al.\citep{nie2015effect} analyzed the impact of degree correlation on controllability; and Menichetti et al.\citep{Menichetti2015} addressed the robustness and stability of control configuration. 

We can already build on a number of works that address the role of interconnecting links between different network layers ~\cite{Buldyrev2010,Gomez2013,Baxter2012a,Cardillo2013}. 
It was shown recently that structural network properties can change even in an abrupt way~\cite{Radicchi2013}. Furthermore, interconnecting links can significantly affect the way \emph{dynamic processes} evolve in multilayer networks~\cite{Aguirre2014,Wang2013a,Garas2014a,Burkholz2015}. 

However, even in the simplest cases, knowing the dynamics on a multilayer network does not mean that we also can control it, i.e. steer the dynamics toward a desired final state. 
This problem can be re-casted as a {\it design} problem: Given a network with two layers, how can we connect these layers with a limited number of inter-layer links such that the whole network can be controlled by using driver nodes from only one layer? 
If not all the nodes in the network can be controlled, what is  the size of the controllable subnetwork that can be controlled by a fixed number of driver nodes. 
In our work, which can be considered as a proof-of-concept, using extensive computer simulations to test different driver node selection criteria and four distinct network interconnecting strategies, we demonstrate that 
a) the whole two-layer network can be controlled by driver nodes from just one layer,
b) to maximize the controllable subnetwork, \emph{peripheral} nodes should be used to connect the two layers 
c) choosing peripheral nodes to control the network can be as efficient as choosing central nodes. 

\section{Model description}
We consider a two-layer network $G(V,E)$ with the number of nodes $N=\abs{V}=N_{0}+N_{1}$, where layer 0 contains $N_{0}$ and layer 1 $N_{1}$ nodes. 
The links in each layer are \emph{directed}, i.e. each node $i$ has an in-degree $k^{\mathrm{in}}_{i}$ and an out-degree $k^{\mathrm{out}}_{i}$.
Building on Ref~\cite{Menichetti2014}, both are drawn independently from a  power-law degree distribution $P(k)\propto k^{-\gamma}$ with $k_{\mathrm{min}}=1$ and $\gamma=2$, using the uncorrelated configuration model (Results for different power-law exponent are shown in supporting information)~\citep{Catanzaro2005}.
We can combine these degree values to assign an importance value $w_{i}$ to each node
\begin{equation} 
\label{eq2} 
w_i=\left(k_{i}^{\mathrm{in}}\right)^{\alpha} \ \left(k_{i}^{out}\right)^{(1-\alpha)}.       
\end{equation}
Here $\alpha$ is a free parameter ranging from 0 to 1.
As $\alpha$ increases, more importance is attributed to the in-degree in the calculation of $w$.
We refer to \emph{central nodes} as nodes with a high importance value $w$.  

Layers 0 and 1 are connected by $L$ additional bidirectional interlayer links (We also test the case for interlayer links of randomly assigned direction, as shown in supporting information). 
Only one interlayer link per node is allowed at maximum, with $q=L/N_{0}$ as the fraction of interlayer links.

Our main assumption is that we can only access the $N_{0}$ nodes on layer 0, in order to control the $N$ nodes in the whole network. 
If we can control $N_{c}\leq N_{0}$ nodes of layer 0, what is the number $N_{b}\leq N$ of nodes in the whole network that we can  control, directly or indirectly, by means of $N_{c}$?
Ideally, we want to choose $N_{c}$ as small as possible, while $N_{b}$ reaches values close to $N$. 

The choice of this ideal $N_{c}$ of course also depends on the strategy by which the two layers are coupled using the $q$ interlayer links. 
Hence, our main research question is \emph{how to couple} these two layers in order to maximize $N_{b}$. 
Given the scale-free degree distribution, for each layer we can distinguish between \emph{hubs}, i.e. nodes with a high importance value $w_{i}$, and nodes with low $w_{i}$ that are only loosely integrated in the layer network. We refer to the latter as \emph{periphery}.
 
\begin{figure}[t]
  \centering
  \includegraphics[width=0.7\columnwidth]{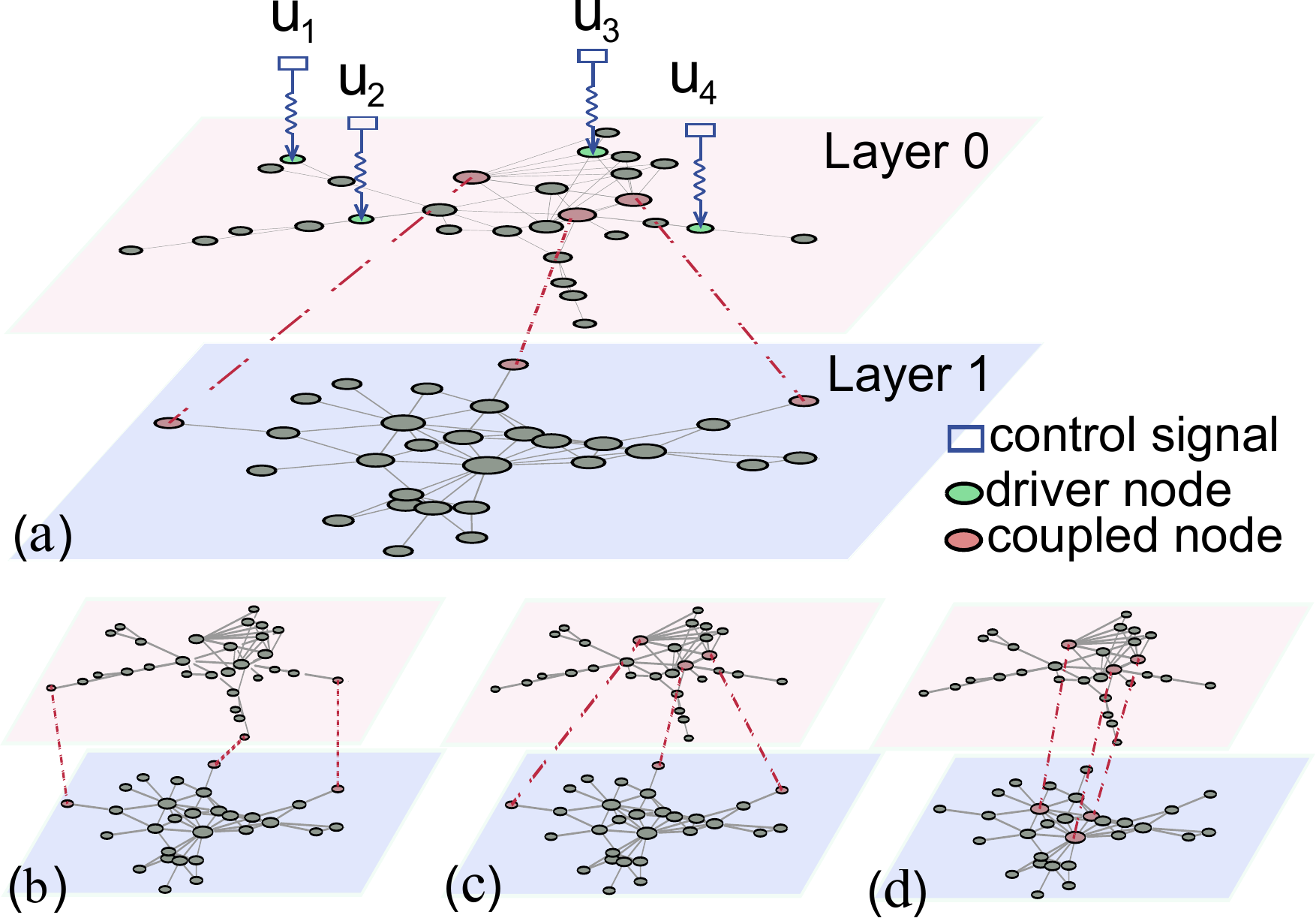}
  \caption{(Color online) Illustration of the coupling strategies for two layer networks. (a) The main scenario where we aim to control the system of networks using only driver nodes from layer 0.  (b) The PP interconnecting strategy. (c) The CP interconnecting strategy. (d) The CC interconnecting strategy. Nodes colored in red are coupled by interlayer links denoted by red dashed lines. In the illustration, the two layer networks are interconnected by $L=3$ links.}
  \label{figure1}
\end{figure}

For the connection of the two layers, we can now think of four different strategies ~\cite{Aguirre2013}, shown in Fig.~\ref{figure1}: 
(CC) nodes with high importance value $w$ in layer 0 are connected to nodes with high importance value $w$ in layer 1, 
(CP) nodes with high $w$ in layer 0 are connected to peripheral nodes with low $w$ in layer 1,
(PC) peripheral nodes in layer 0 are connected to  nodes with high $w$ in layer 1, and
(PP) peripheral nodes in layer 0 are connected to peripheral nodes in layer 1. 

The strategy to couple the two layers now consists of two steps: (i) calculate $w_{i}$ and rank the nodes with respect to their $w_{i}$, for each of the two layers separately.
(ii) until $q$ is reached, deterministically choose nodes according to their rank on each of the two layers. 
I.e. dependent on the different strategies, we link high or low ranked nodes from the two layers until $L$ interlayer links are formed. 

\section{Structural Controllability}
In order to apply the framework of structural controllability \citep{Liu2011}, we need to make assumptions about the dynamics that change intrinsic properties of the nodes. 
Let us assume that each node is characterized by a variable $x_{i}(t)$. 
As shown in Figure \ref{figure1}a, some of these nodes can be influenced by external signals $u_{k}(t)$, which shall later be used to control the dynamics of the whole network.
Let $\mathbf{U}(t)  \in \mathbb{R}^{N_{c}}$ be the vector of control signals. Then the matrix $\mathbf{B} \in \mathbb{R}^{N \times N_c}$ defines which nodes are directly controlled by the external signals $u_{k}(t)$ $(k=1,..,N_{c})$, with the element $b_{ij} \neq 0$ if signal $j$ is attatched to node $i$. 
  
The framework of structural controllability requires us to choose a linear dynamics for $x_{i}(t)$, which reads in vector notation $\mathbf{X}(t)=\{x_{1}(t),x_{2}(t),...,x_{N}(t)\}$ with
$\mathbf{X} \in \mathbb{R}^N$:
\begin{equation} \label{eq1}
\dot{\mathbf{X}}(t)=\mathbf{AX}(t)+\mathbf{BU}(t), 
\end{equation}
$\mathbf{A} \in \mathbb{R}^{N \times N}$ is the interaction matrix with elements $a_{ij}$ $(i,j=1,...,N)$ that describe the weighed influence between any two nodes either within or across layers in the multilayer network.
According to the Kalman rank condition~\cite{kalman1963mathematical}, the dynamical system defined by \eqref{eq1} is controllable, i.e. it can be driven from an initial state to any desired state, if and only if the controllability matrix $\mathbf{C}=[\mathbf{B,AB,A^2B,...,A^{N-1}B}] \in \mathbb{R}^{N \times (N\cdot N_c)}$ has full rank, i.e., rank$(\mathbf{C})$=$N$. 

In some cases, the exact value of the nonzero elements in $\mathbf{A}$ and $\mathbf{B}$ is not available, and the precise computation of rank($\mathbf{C}$) is therefore unattainable.
For those cases, the weaker requirement of structural controllability~\cite{Lin1974} was introduced. 
It treats  $\mathbf{A}$ and $\mathbf{B}$ as structured matrixes, i.e. their elements are either fixed zeros or free parameters.  

The system is structurally controllable if the maximum rank of $\mathbf{C}$, denoted as $rank_g(\mathbf{C})$, can reach $N$ as a function of the free parameters in $\mathbf{A}$ and $\mathbf{B}$.
Based on \citep{Lin1974}, \citet{Liu2011} derived the minimum input theorem to identify the \emph{minimum} number of driver nodes $N_d$ needed  to control the \emph{whole} network of $N$ nodes. 

In real situations, some driver nodes necessary for control may not be accessible or the number of driver nodes $N_{d}$ may be too large to be efficiently influenced by the limited number of control signals. 
Hence full control of the network cannot be achieved.
In those scenarios, we are interested in the size of the subnetwork, given by $N_{b}=rank_g(\mathbf{C}) \leq N$, that can still be controlled by a given set of driver nodes, $N_{c}\leq N_{d}$.

If there is only one driver node $i$ denoted by $\mathbf{B}$, the value of $N_b$ defines the control centrality of $i$~\cite{Liu2012,Pan2014b}. 
For more than one driver node, there is an overlap in the  subspace controlled by each node and the sum of the control centralities of all driver nodes may overestimate $N_b$.

Despite that analytical treatment using tools from statistical physics can be found in literature for the case of $N_{d}$ when the controllability of the whole system is considered, to the best of our knowledge, there is no analytical method to predict $N_{b}$ when we focus on the controllable subsystem. Even for the simplest case where only a single driver node is considered, and for very particular topologies, such as a directed acyclic graph in which a unique hierarchical structure can be identified, can the controllable space size of one node be predicted by its hierarchical level.
Therefore, for a general case with more than one driver node, in order to effectively determine $N_b$, we deploy a linear programming approach \cite{Poljak1990}.

The algorithm works by constructing an auxiliary network that is larger than the original one, to identify the cycle partition structure \cite{Poljak1990} that contains the maximum number $N_{b}$ of  controllable nodes in the complex network. 
We first construct an initial auxiliary network $H(\hat{E},\hat{V})$ that contains $N_{H}=\hat{\abs{V}}=N+N_{c}$ nodes.
$N_{c}$ is the number of control signals, which are represented by an additional set $S_{c}$ of nodes in the auxiliary network. 
Regarding its topology, $H$ preserves all the links defined in the matrix $\mathbf{A}$, but has additional links from the set $S_{c}$ to the driver nodes $N_{c}$ (one link per driver node, indicated in Fig. \ref{figure1} by the zig-zag arrows).
Next, we identify the reachable network $H^{\prime}$ which is given by those nodes that can be reached via a directed path from the set $S_{c}$. 
Then, we change the topology of $H^{\prime}$ by adding directed links of weight zero from any node within $H^{\prime}$ (excluding the set $S_{c}$) to all nodes in the set $S_{c}$. 
Also, we add self-loops of weight zero to all the nodes in $H^{\prime}$, to arrive at the auxiliary network $H^{\prime}(\tilde{E},\tilde{V})$ 

From $H^{\prime}$, we can now calculate $N_{b}$  as the optimal value of the integer linear problem
\begin{equation} \label{eq3}
\max \sum\nolimits_{e \in \tilde{E}} w_e h_e
\end{equation}
where $w_e$ denotes the weight of link $e$, and $h_{e}\in \{0,1\}$ is a binary variable indicating whether one link is chosen to be part of the optimal solution of the linear programming problem. 
The subjections
\begin{equation} 
\label{eq4}
\sum_{e \ {\rm leaves} \ v} h_e=1;\; \sum_{e  \ {\rm enters} \ v} h_e=1  \ \forall  v \in \tilde{V}
\end{equation}
guarantee that the optimal solution to Eq.\eqref{eq3} forms a cycle partition that spans the graph $H^{\prime}$.

\begin{figure}[t]
  \centering
  \includegraphics[width=0.7\columnwidth]{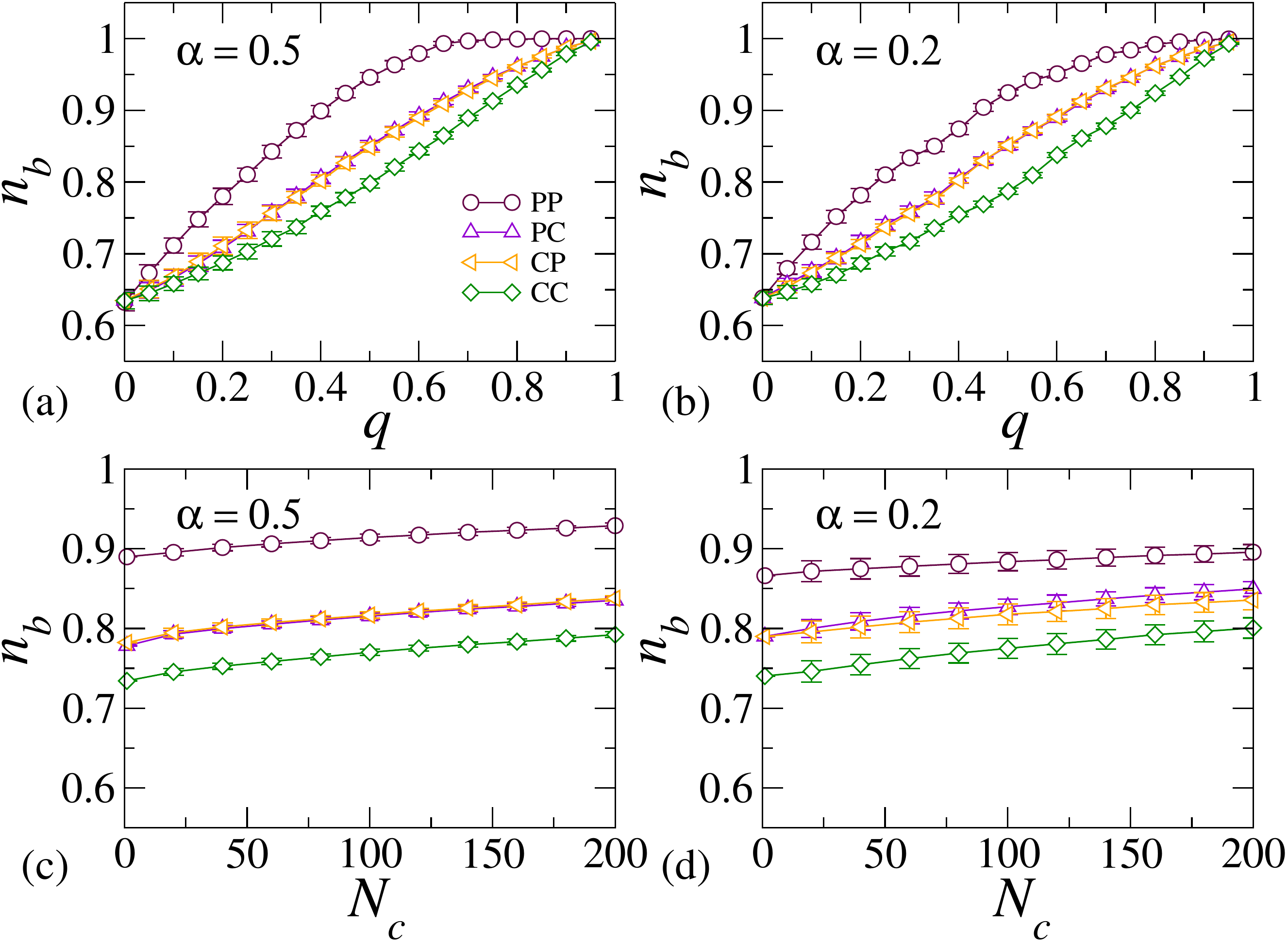}
  \caption{(Color online) $n_b$ as a function of the fraction of interlayer links and the number of driver nodes. The top panel shows the characteristics of $n_b$ as a function of the fraction of interlayer links $q$. In (a) $\alpha=0.5$, in (b) $\alpha=0.2$.The results are produced with $30$ driver nodes. The bottom panel shows the characteristics of $n_b$ as a function of the number of driver nodes $N_c$, for $q=0.4$. In (c) $\alpha=0.5$, in (d) $\alpha=0.2$. Change of the number of driver nodes to other values smaller than $N_d$ doesn't alter our findings. All the results are produced with $N_{0}=N_{1}=2000$. The data points are obtained over 100 simulations, with error bars representing standard deviation.}
  \label{figure2}
\end{figure}

\section{Results and discussions}
We now apply the above optimization method to treat the multilayer networks that were constructed by the four different coupling strategies. 
To obtain the results, we use an ensemble approach, i.e. we keep the configuration of each layer constant, but generate  5 multilayer network realizations with $N_0=N_1=2000$ for each possible parameter configuration and for each coupling strategy.  The importance parameter $\alpha$ and the fraction of interlayer links $q$ are both varied between 0 and 1 in steps of 0.05.
Eventually, for each configuration of parameters and each coupling strategy, we randomly sample $100$ sets of driver nodes from layer 0, i.e. 20 per multilayer network realization. 
This results in $1.6\times10^5$ different network configurations in total. 

\begin{figure}[t]
  \centering
  \includegraphics[width=0.7\columnwidth]{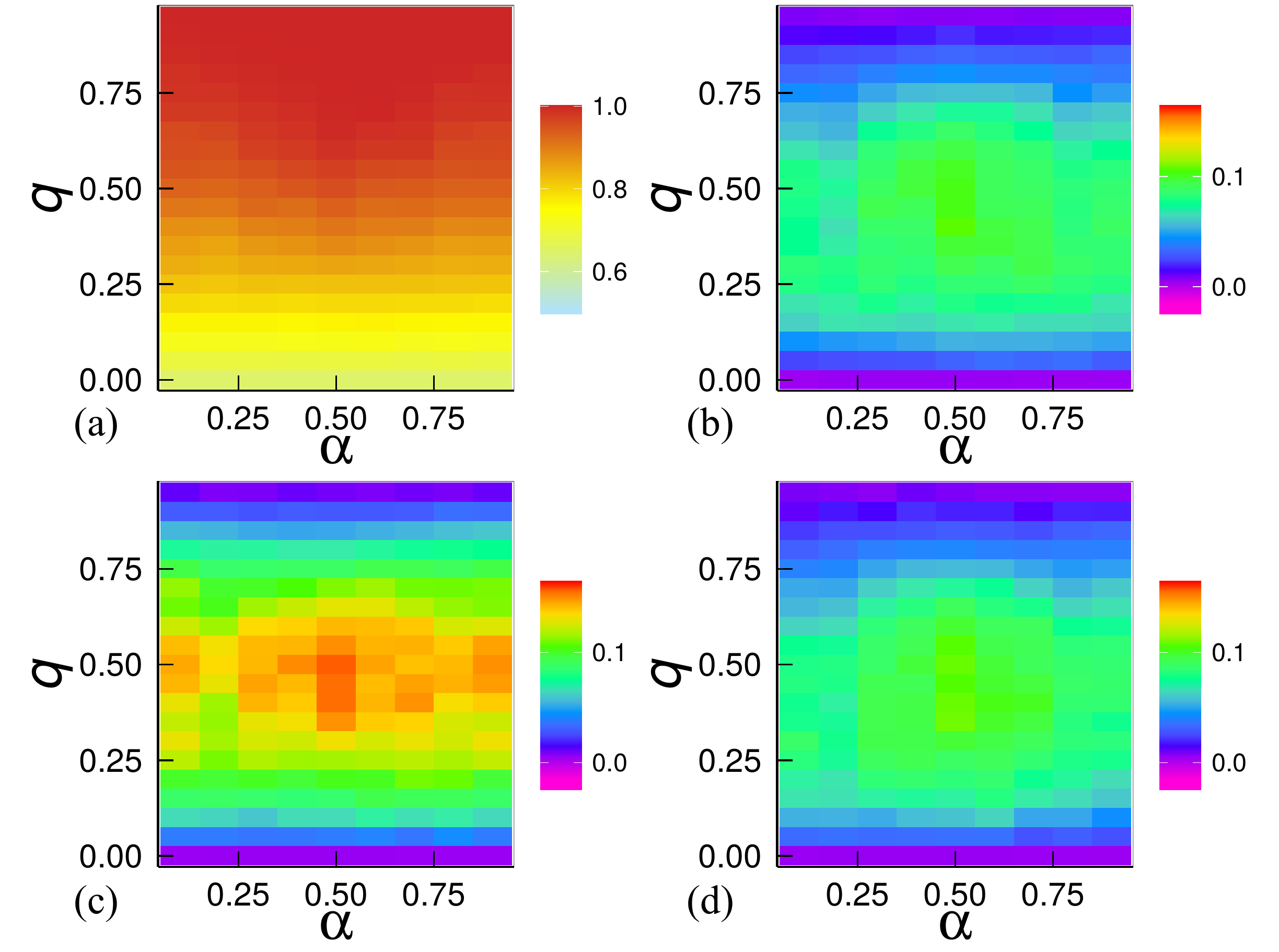}
  \caption{(Color online)(a) Color map encoding the fraction of controlled network $n_b$ on the  $\alpha-q$ parameter plane for PP strategy. (b-d) Color map encoding the difference between $n_b$ for PP and $n_b$ for CP,CC,PC strategies. A positive value indicates that $n_b$ for PP is greater than the corresponding strategy. The above results are obtained with $N_c=30$ driver nodes.}
  \label{figure3}
\end{figure}

Our main interest is in the relative size of the network that can be controlled this way, i.e. we calculate $n_{b}=\mean{N_{b}}/N$ for each parameter configuration and coupling strategy. 
$n_b=1$ indicates that for this configuration the whole system can be controlled.   
Our results are presented in two different ways: In the top panel of Fig.~\ref{figure2}, we compare $n_b$ dependent on the fraction of interlayer links $q$, with the number of driver nodes $N_{c}$ kept constant, while in the bottom panel of Fig.~\ref{figure2}, $n_b$ is shown dependent on the number of driver nodes $N_{c}$, with $q$ kept constant. 
The results are presented for two values of the importance parameter $\alpha$, but results for varying $\alpha$ are shown in Fig. \ref{figure3}.

From the top panel of Fig.~\ref{figure2}, we report two observations: (i) For the PP strategy, a sizable control of the multilayer network (i.e. $n_{b}$ reaches values close to $1$) can be reached already for a fraction of interlayer links $q$ below 1, i.e. not every node in the two layers need to be linked. E.g., for $q=0.5$, $n_{b}$ already ranges between 0.8 and 1. (ii) Among the four different strategies to couple the two layers, the PP strategy fares best. I.e. with respect to control, coupling peripheral nodes is more beneficial than coupling nodes with high importance value. The results are similar for the two $\alpha$ values chosen.   

From the bottom panel of Fig.~\ref{figure2}, we observe again two interesting findings for this particular system: (i) the range of the controllable network, $n_{b}$, does not strongly vary if we increase the number of driver nodes $N_{c}$ from one to a value much smaller than the system size, such as 200 as shown in the figure. $n_{b}$ never reaches 1, given the small values of $N_{c}\ll 2000$, but already reaches remarkable values between $0.7$ and $0.95$ (Results for larger values of $N_{c}$ are reported in the supporting figure Fig.S3(a)). 
(ii) Again, the PP strategy to link peripheral nodes in both layers allows of a considerably better control of the network. This distinction becomes most pronounced for $\alpha=0.5$. 

The superiority of the PP strategy in connecting the two layers is further demonstrated in Fig.~\ref{figure3}, where we explore the full parameter space of $\alpha$ and $q$.
Fig.~\ref{figure3}(a) shows, for the PP strategy, the gradual increase in $n_{b}$ as the fraction $q$ of interlayer links increases. 
There is no strong dependency on the importance parameter $\alpha$, only a slight improvement for $\alpha=0.5$ which weights the in-degree and the out-degree equally.  
Fig.~\ref{figure3}(b-d) illustrate the difference between the PP strategy (a) and the remaining other strategies, by just plotting the difference in $n_{b}$ compared to Fig.~\ref{figure3}(a). This difference is always positive, indicating the advantage of the PP strategy.
More remarkable, the difference becomes the largest for moderate values of $q$ and $\alpha$. 
Obviously, for $q$ close to $0$ or $q$ close to $1$, the four strategies become indistinguishable.

\begin{figure}[t]
  \centering
  \includegraphics[width=0.7\columnwidth]{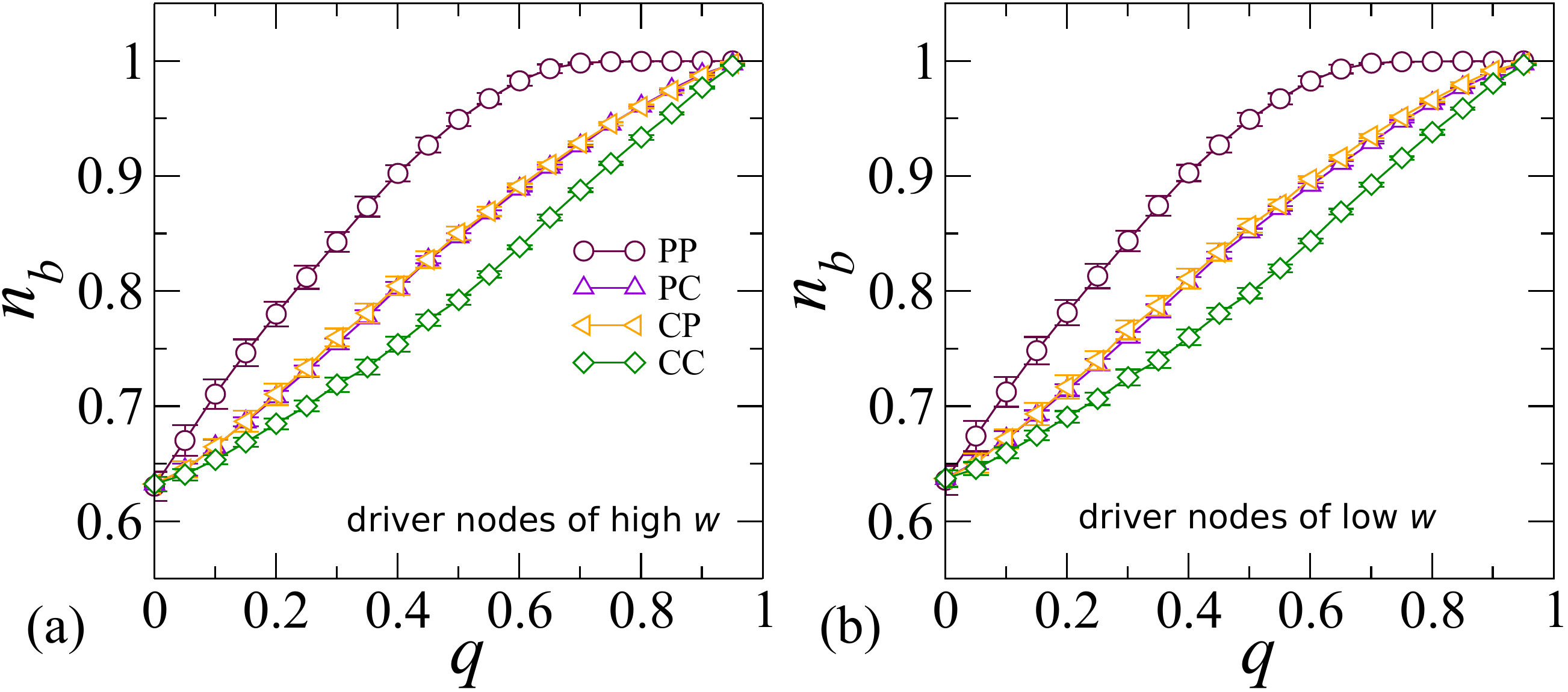}
  \caption{(Color online) $n_b$ as a function of the fraction of interlayer links. In (a) driver nodes are sampled from nodes of high $w$. In (b) driver nodes are sampled from nodes of low $w$. The results are produced with $N_c=30$ driver nodes.  }
  \label{figure4}
\end{figure}

So far, we have only investigated the impact of different strategies in \emph{linking} the two layers. 
But we did not discuss whether it is more beneficial to choose central nodes or peripheral nodes from layer 0 as \emph{driver nodes}. 
In fact, to obtain Figs. \ref{figure2}, \ref{figure3} we have randomly sampled the set of driver nodes from layer 0. 
Therefore, in Fig. ~\ref{figure4} we now investigate the impact of central or peripheral \emph{driver} nodes on the controllability of the multilayer network.
Specifically, we compare two scenarios:  in Fig.~\ref{figure4}(a) the driver nodes are sampled from the top $10\%$ nodes of high $w$ in layer 0, whereas in 
Fig.~\ref{figure4}(b) the driver nodes sampled from the top $10\%$ nodes of low $w$ values in layer 0.  
A comparison of the four discussed strategies to connect the two layers and all values of $q$, shows that the difference in $n_b$  can be almost neglected (it is less than 1\% as shown in the supporting figure Fig.S3(b)). 
This indicates that, by injecting control signals into \emph{peripheral nodes}, we can control as much of the total network as by injecting control signals into \emph{central nodes}. 

\section{Conclusions }
In this paper, we study the controllability of two-layer directed networks with numerical simulations. In our model, we distinguish between two different kind of nodes: (i) the nodes that should be chosen to \emph{connect} the two layers, in order to maximize the number of controllable nodes in the whole network, $N_{b}$.
(ii) the driver nodes that should be chosen on layer 0 to \emph{control} this subspace. 
These nodes do not necessarily have to be the same. 
The number of interlayer connections is determined by the parameter $q=L/N_{0}$, whereas the number of driver nodes can vary as well, $N_{c}\leq N_{0}$. 
For a given $N_{c}$, increasing $q$ usually leads to increasing $N_{b}$, until $N_{b}$ reaches its saturation.

At the same time, for given $q$ and $N_{c}$, there is a preferred coupling strategy to maximize $N_{b}$, which is coupling \emph{peripheral} nodes in \emph{both} layers. 
Assuming that it is less costly to access peripheral nodes as compared to central nodes, the PP strategy would also be the most cost-efficient strategy.
We emphasize that this finding differs from \citep{Aguirre2014} where the CC strategy was preferred. 
This was found for an \emph{undirected} network, on which a synchronization dynamics was investigated, whereas we consider a \emph{directed} network, on which we  assume a linear dynamics. 

As a second important finding, we have shown that the control of the network can be as \emph{effectively} achieved by choosing $N_{c}$ from the \emph{peripheral} nodes as from the central 
nodes. Referring to the cost argument above, choosing peripheral nodes as driver nodes is both effective and cost-efficient. 

The third finding points to the size of the controllable subspace $N_{b}$. Here, we show that it is sufficient to choose driver nodes from just one layer, to control the whole two-layer network, in accordance with earlier findings \cite{Liu2011}.
Dependent on the fraction of interlayer links, the full control can be even achieved with a small number of driver nodes, $N_{c}\ll N_{0}$ (e.g. $N_{c}=30$ for $N_{0}=2000$ and $q=0.6$), given that the most efficient PP strategy is used for linking the layers. 
This again emphasizes the importance of peripheral nodes in controlling multilayer networks. 

The above findings were obtained for the particular system of networks used in our simulations. However, we also checked the robustness of our findings with the alternative configurations presented in the supporting information.
These configurations include the case with randomly assigned directionality for the interlayer links, and the case of scale free networks with another network exponent. 
This indicates that peripheral nodes may play similarly important role in connecting and controlling multilayer networks for systems with other network configurations as well. 
In this sense, our results could be useful for practical applications. 
As an example let us consider the coupling between a power grid and a communication network. A regulator may want to achieve better controllability of the full system by accessing a small number of driver nodes from the communication network, and this could be achieved by identifying optimal ways to couple the two networks using a methodology similar to the one presented in our paper.
However, a more concrete exploration of the way our approach can be used for real systems is beyond the scope of the current manuscript, and is left for future studies.

\paragraph{Acknowledgements.}
We gratefully acknowledge helpful discussions with Y.Y Liu from Harvard Medical School. A.G. and F.S. acknowledge financial support by the EU-FET project MULTIPLEX 317532.

\section*{Appendix}

We now test the robustness of the our main results by considering two alternative configurations of our model:
\begin{enumerate}
 \item[a)] We build two-layer networks using a power law degree distribution with $\gamma$ =2.5.
 \item[b)] We connect two layers of networks by interlayer links with directions assigned randomly. 
\end{enumerate}
In Figures S1 and S2 we show results obtained with the above alternative configurations using networks of size $N_{0}=N_{1}=2000$. 
The presented results are averaged over 100 simulations with error bars representing the standard deviation. 
Even though some effects are less pronounced (for example, see Fig.S1 wrt Fig.2), the results are in-line with our main findings.

\setcounter{figure}{0}
\makeatletter 
\renewcommand{\thefigure}{S\@arabic\c@figure}
\makeatother

\begin{figure}[h]
  \centering
  \includegraphics[width=0.65\columnwidth]{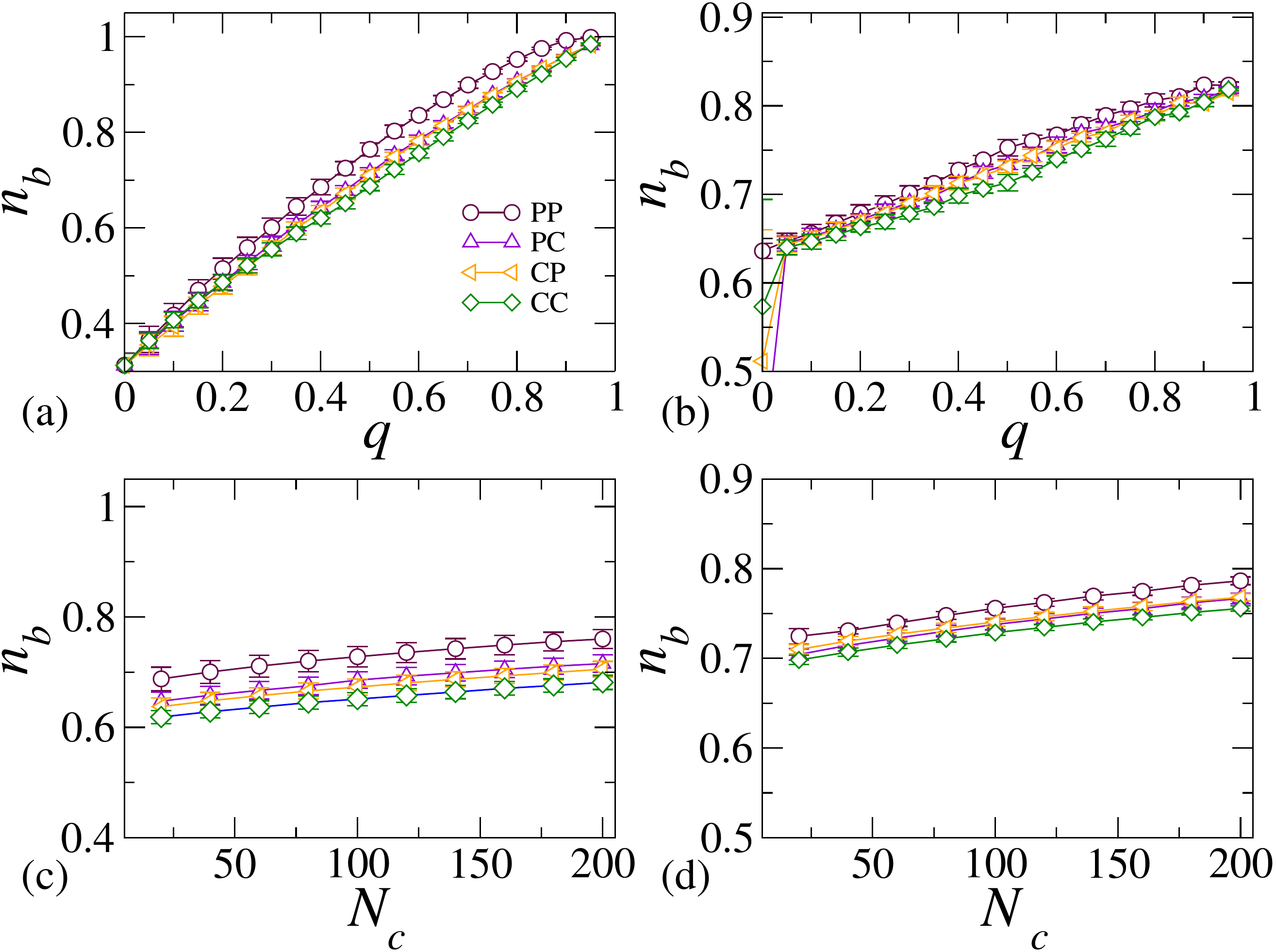}
  \caption{(Color online) $n_b$ as a function of the fraction of interlayer links and the number of driver nodes. The top panel shows the characteristics of $n_b$ as a function of the fraction of interlayer links $q$. The bottom panel shows the characteristics of $n_b$ as a function of the number of driver nodes $N_c$, for $L=800$ interconnecting links. In (a)\&(b) $\gamma$ =2.5, and the two network layers are connected by bidirectional interconnecting links. In (c)\&(d) $\gamma$ =2.5, and the two network layers are connected by interconnecting links whose directions were assigned randomly. The results are produced with $30$ driver nodes and $\alpha=0.5$, and are consistent with the results shown in Fig.~2.}.
\end{figure}

\begin{figure}[b]
  \centering
  \includegraphics[width=0.6\columnwidth]{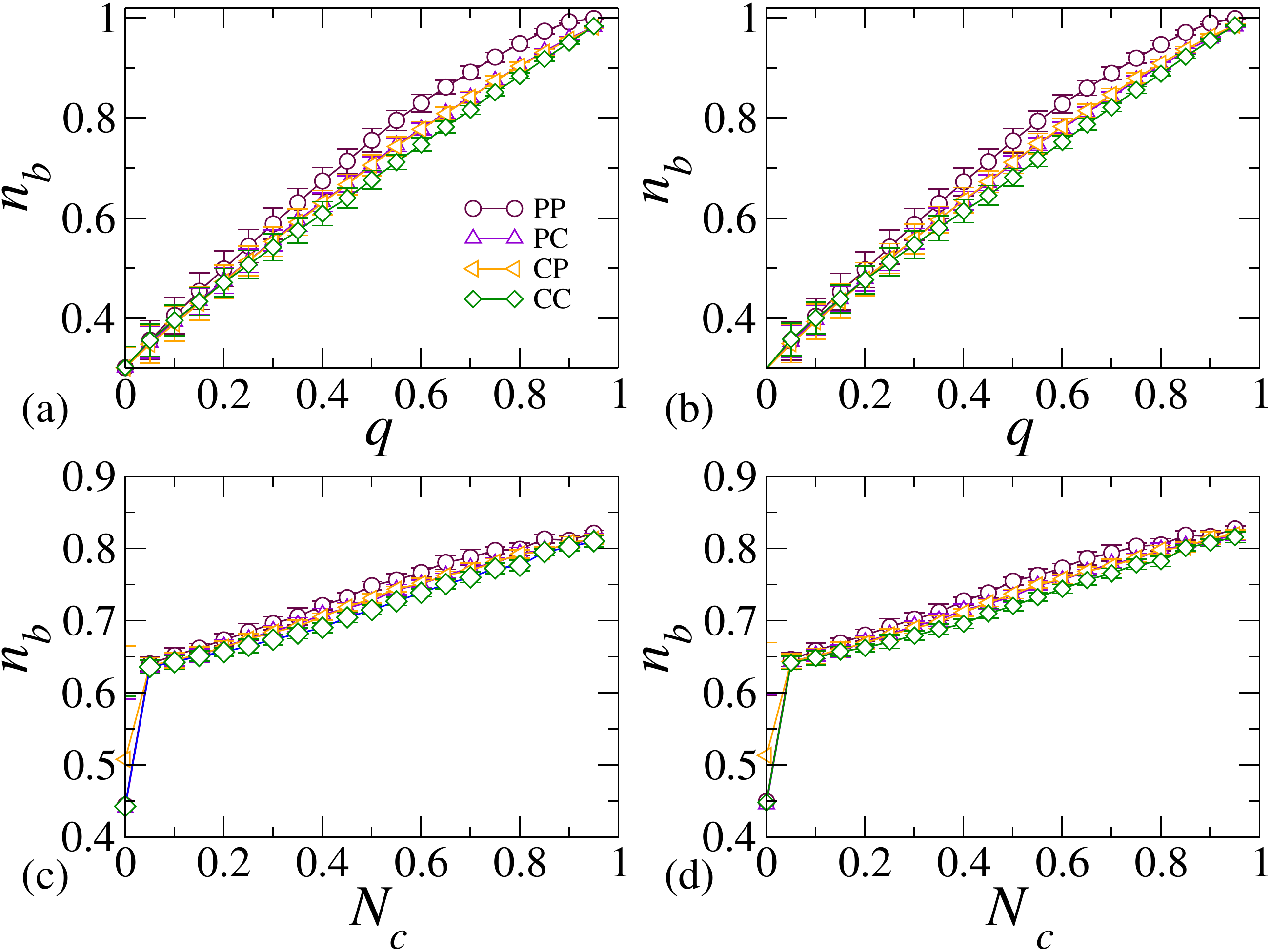}
  \caption{(Color online) $n_b$ as a function of the fraction of interlayer links. The top two figures are obtained with $\gamma$ =2.5, and two layers of networks are connected by bidirectional interconnecting links. The bottom figures are produced with $\gamma$ =2.0, and two layers are connected by  interconnecting links whose directions were assigned randomly. In (a)\&(c) the driver nodes are sampled from nodes of high $w$, while in (b)\&(d) the driver nodes are sampled from nodes of low $w$. All the results are produced with $N_c=30$ driver nodes, and are consistent with the results shown in Fig.~4.}.
\end{figure}

\begin{figure}[t]
  \centering
  \includegraphics[width=0.6\columnwidth]{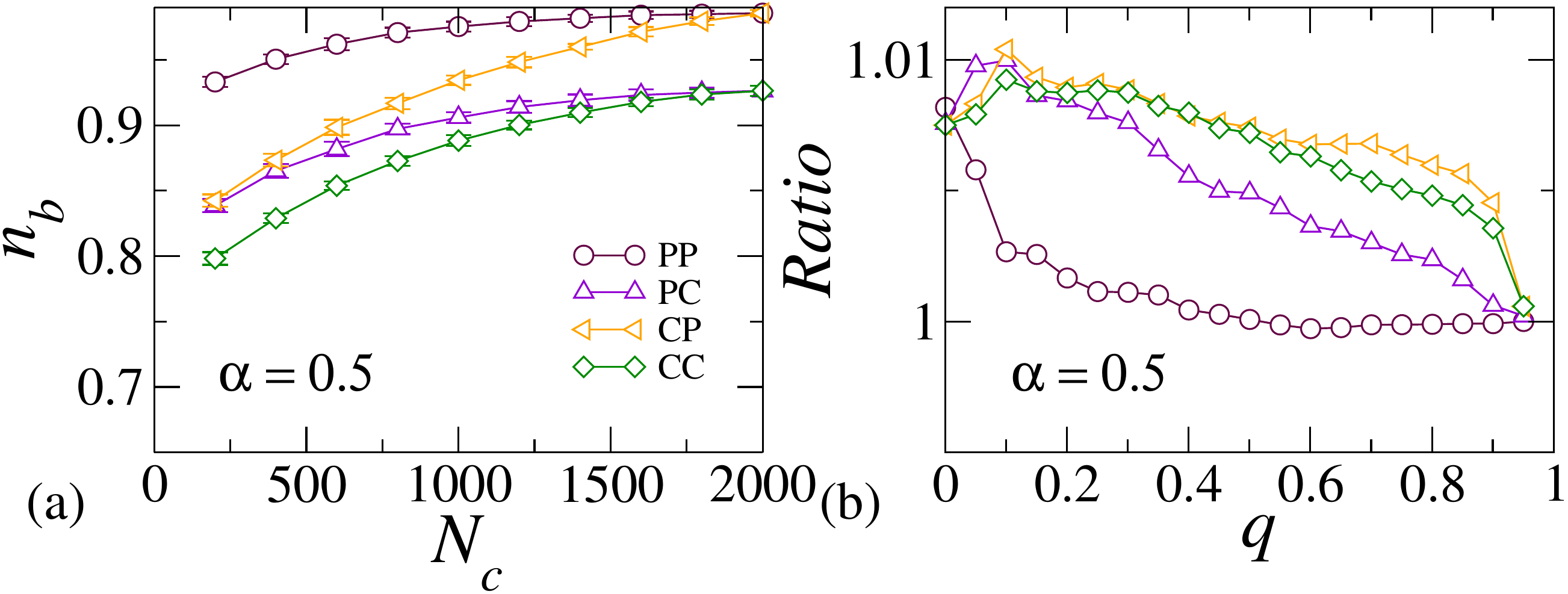}
  \caption{(Color online) $n_b$ as a function of the number of driver nodes $N_c$, for $q=0.4$. Increasing the number of driver nodes $N_c$ to $N_0=2000$ confirms that the PP strategy to link peripheral nodes in both layers allows to better control the whole network. The data points are obtained over 100 simulations, with error bars representing standard deviation. (b) The ratio of average $n_b$ with driver nodes of low $w$ over the average $n_b$ with driver nodes of high $w$. This sub-figure shows that the difference in $n_b$ is marginal and can be safely neglected. The figure is produced using the two-layer network configuration that is discussed in the main text.}
\end{figure}

\clearpage

\bibliographystyle{apsrev4-1}
\bibliography{ZGS-mc}
\end{document}